\documentclass[12pt,dvips]{article}

\usepackage{rotating}
\usepackage{axodraw}
\usepackage{epsfig}
\def\lsim{\mathrel{\raise.3ex\hbox{$<$\kern-.75em\lower1ex\hbox{$\sim$}}}}
\def\gsim{\mathrel{\raise.3ex\hbox{$>$\kern-.75em\lower1ex\hbox{$\sim$}}}}

\newcommand{\beq}{\begin{eqnarray}}
\newcommand{\eeq}{\end{eqnarray}}

\begin{document}


\mbox{ }\hfill DESY 04-077\\
\mbox{ }\hfill IFT--04/10\\
\mbox{ }\hfill MC--TH--2004--02\\
\mbox{ }\hfill PSI-PR-04-06\\
\mbox{ }\hfill hep--ph/0404119\\


\bigskip
\begin{center}
{\Large{\bf Determining  \boldmath{$\tan\beta$} in $\tau\tau$ Fusion to SUSY 
            Higgs\\[2mm] Bosons at a Photon Collider}}\\[1cm]
S.~Y.~Choi$^1$, J.~Kalinowski$^2$, J.S.~Lee$^3$,
M.M.~M\"uhlleitner$^4$, M.~Spira$^4$\\[2mm] 
and P.~M.~Zerwas$^5$
\end{center}

\bigskip

\begin{enumerate}
\item[{}] $^1$ Dept.\ Physics, Chonbuk National University, 
               Chonju 561--756, Korea
\item[{}] $^2$ Inst.\ Theor.\ Physics, Warsaw University, PL--00681 Warsaw, 
               Poland
\item[{}] $^3$ Dept.\ Physics and Astronomy, Univ.\  Manchester, 
               Manchester M13 9PL, UK
\item[{}] $^4$ Paul Scherrer Institut, CH-5232 Villigen PSI, Switzerland
\item[{}] $^5$ Deutsches Elektronen--Synchrotron DESY, D--22603 Hamburg, 
Germany
\end{enumerate}
\bigskip
\bigskip\bigskip

\begin{abstract}
We investigate $\tau\tau$ fusion to light $h$ and heavy $H$ and $A$ 
Higgs bosons in the Minimal Supersymmetric Standard Model (MSSM) at a photon 
collider as a promising channel for measuring large values of $\tan\beta$. 
For standard design parameters of a photon collider an error 
$\Delta\tan\beta\sim 1$, uniformly for $\tan\beta \gsim 10$, may be expected, 
improving on  complementary measurements at the LHC and $e^+e^-$ linear 
colliders.

\end{abstract}
%



\renewcommand{\thefootnote}{\alph{footnote}}
\newpage

\noindent
{\bf 1. \underline{Introduction.}}
The measurement of the mixing parameter $\tan\beta$, one of the fundamental 
parameters in the Higgs sector of the
Minimal Supersymmetric Standard Model (MSSM) and other supersymmetric
scenarios, is a difficult task. Many of the observables, in the
chargino/neutralino sector \cite{R1} for instance, involve only $\cos2 \beta$,
the slope of which approaches $-4/\tan^3\beta$ for large values of $\tan\beta$
and thus are quite insensitive to the parameter $\tan\beta$ in this range.
Remarkably different however are the heavy $H/A$ Higgs couplings to down-type
fermions which, for values of the pseudoscalar Higgs boson mass at the
electroweak scale and beyond, both are directly proportional to $\tan\beta$ if
this parameter becomes large, see {\it e.g.} Ref.~\cite{R2}, and which thus
are highly sensitive to its value. Also the down-type couplings of the light
$h$ Higgs boson in the MSSM are close to $\tan\beta$ if the pseudoscalar mass
is moderately small.

The Higgs down-type fermion vertices play a decisive role in various
observables which can be exploited for measuring $\tan\beta$. Examples are the
total widths of the $H/A$ Higgs bosons \cite{R3},  Higgs-bremsstrahlung off
bottom quarks in $e^+e^-$ collisions \cite{R4}, and polarization effects
\cite{R4a}.  A summary of the expected
results has been presented recently in Ref.~\cite{R5}. Moreover, $b$-quark
fusion to $H/A$ bosons at the LHC has recently been investigated in a detailed
experimental simulation \cite{R6} and proved quite promising for large values
of $\tan\beta$. All these methods are applicable in part of the MSSM parameter
space and expected accuracies generally hover around the 10\% level. In this
situation any additional method for measuring $\tan\beta$ is valuable,
improving the picture significantly even if the individual error remains of
similar size.

In this note we point out that $\tau\tau$ fusion to Higgs bosons at a photon
collider can provide a valuable method for measuring $\tan\beta$ -- a natural
next step after searching for Higgs bosons in $\gamma\gamma$ fusion
\cite{R6a} and exploiting the associated loop-induced production cross section
\cite{C7a}. The entire Higgs mass range can be covered by this method up to 
the kinematical limit for large $\tan\beta$.

In the next section a semi-quantitative analytical discussion of $\tau \tau$
fusion to Higgs bosons will motivate the detailed study. The fusion process is
the dominant Higgs boson production process in $\gamma\gamma$ collisions
giving rise to $\tau^+\tau^-$ + Higgs in the final state. In the subsequent
numerical analysis, however, a full set of signal and background processes is
taken into account. We demonstrate that for standard design parameters of a
photon collider an error $\Delta\tan\beta\sim 1$ in measurements of a large
$\tan\beta$ value can be expected. \\

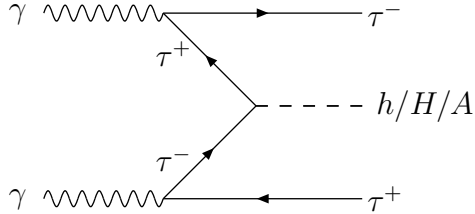
\begin{figure}[htb]
\begin{picture}(700,100)(100,0)
\Text(221,85)[]{$\gamma$}
\Photon(230,85)(275,85){3}{7}
\Text(221,15)[]{$\gamma$}
\Photon(230,15)(275,15){3}{7}
\ArrowLine(310,50)(275,85)
\ArrowLine(275,15)(310,50)
\Text(375,50)[]{$h/H/A$}
\ArrowLine(275,85)(350,85)
\Text(360,85)[]{$\tau^-$}
\Text(280,30)[]{$\tau^-$}
\ArrowLine(350,15)(275,15)
\Text(360,15)[]{$\tau^+$}
\Text(280,70)[]{$\tau^+$}
\DashLine(310,49.96)(350,49.96){5}
\DashLine(310,50.04)(350,50.04){5}
\end{picture}
\caption{\it The process of $\tau\tau$ fusion  to Higgs bosons in 
 $\gamma\gamma$ collisions. \label{fig:process}}
\end{figure}

\vskip 3mm
\noindent 
{\bf 2. \underline{\boldmath{$\tau\tau$} Fusion at a Photon Collider.}} 
The alternative channel to the methods summarized above, $\tau\tau$ fusion to 
the light and heavy $h/H/A$ Higgs bosons at a photon collider, 
{\it cf.}\ Fig.\ref{fig:process}, is based on the two-step process:
\begin{eqnarray}
    \gamma+\gamma \to (\tau^+ \tau^-)+(\tau^+ \tau^-) \to
    \tau^+ \tau^- + h/H/A              \label{eq:process}
\end{eqnarray}
For the large-$\tan\beta$ case studied here in detail as an example -- the
Higgs-mass slope crossing the Snowmass point SPS1b \cite{R11} -- the decay
channels of Higgs bosons to supersymmetric particles are nearly shut and all
the Higgs bosons $\Phi=h/H/A$ decay almost exclusively, {\it i.e.} 80 to 90\%,
to a pair of $b$ quarks. Therefore the final state consists of a pair of
$\tau$'s and a pair of resonant $b$ quark jets.

Unlike LHC, $\tau\tau$ fusion is superior to $b$-quark fusion at a photon 
collider.
This is apparent from a quick estimate of the charges and couplings involved.
$b$-quark fusion is suppressed by the fractional electric charge $e_b=-1/3$
that cannot be compensated by color nor by the enhancement of the
fermion-Higgs Yukawa coupling. In total, the $b$-channel suppression is of
order $3\, (1/3)^4 (m_b/m_\tau)^2 \sim 0.1$ compared with the tau channel.

\smallskip

Energies in the range of more than one hundred GeV, {\it i.e.} the size of the
Higgs-boson masses, quite naturally suggest the application of the
equivalent-particle approximation to the $\gamma \gamma$ process
(\ref{eq:process}). In this approximation, the process can be decomposed into
two consecutive steps: photon splittings to tau pairs, $\gamma \to \tau^+
\tau^-$, followed by the fusion process of two (almost on-shell) taus to the
Higgs bosons, $\tau^+ \tau^- \to \Phi$, {\it cf.}\ Fig.~\ref{fig:process}.
Hence, the cross section is given by the convolution of the fusion cross
section with the $\tau\tau$ luminosity in the colliding photon beams.

The fusion cross section to the Higgs bosons $\Phi$ may be written
\begin{eqnarray}
   \hat{\sigma}[\tau^+\tau^- \to \Phi;\, \hat{s}] = 
   \frac{\pi m_\tau^2}{2v^2} \, g^2_{\Phi\tau\tau} \, 
   \frac{M_\Phi \Gamma_\Phi/\pi}{(\hat{s} 
   - M_\Phi^2)^2+M_\Phi^2 \Gamma_\Phi^2}
\label{tau2higgs}
\end{eqnarray}
where $\hat{s}$ denotes the $\tau^+\tau^-$ invariant energy squared and $v$ is
the Higgs vacuum expectation value, $v\simeq 246$~GeV. The coupling
$g_{\Phi\tau\tau}$ is normalized to the Standard Model Higgs coupling to a tau
pair, $m_\tau/v$. For large $\tan\beta$, the couplings\footnote{Note that in
  contrast to the $b$-quark couplings, the $\tau$-couplings are not strongly
  renormalized by higher-order corrections.} are given by
\begin{eqnarray}
g_{\Phi\tau\tau}& =& \tan\beta 
{\rm ~~~~~~~~~~ for ~} \Phi=A   \nonumber\\
g_{\Phi\tau\tau}&\simeq& \tan\beta 
{\rm ~~~~~~~~~~ for ~} \Phi=h,H
\end{eqnarray}
if the pseudoscalar mass parameter $M_A$ is sufficiently light in the case of
$h$, and sufficiently heavy in the case of $H$, {\it cf.}\ Ref.~\cite{R2} for
details.  The $\tau\tau$ luminosity can be derived from the $\gamma \to
\tau$ structure function \cite{R7}
\begin{eqnarray}
   D^\tau_\gamma(z) = \frac{\alpha}{2\pi} 
   [z^2+(1-z)^2] \log\frac{\mu_F^2}{m_\tau^2}     \label{taulumi}
\end{eqnarray}
where the energy fraction transferred from $\gamma$ to $\tau$ is denoted by 
$z$ and non-logarithmic terms have been neglected. The factorization scale
$\mu_F$ 
typical for the subsequent fusion process is set by the Higgs mass in 
this calculation, $\mu_F=M_\Phi$. 

\begin{figure}[ht!]
\begin{center}
\epsfig{figure=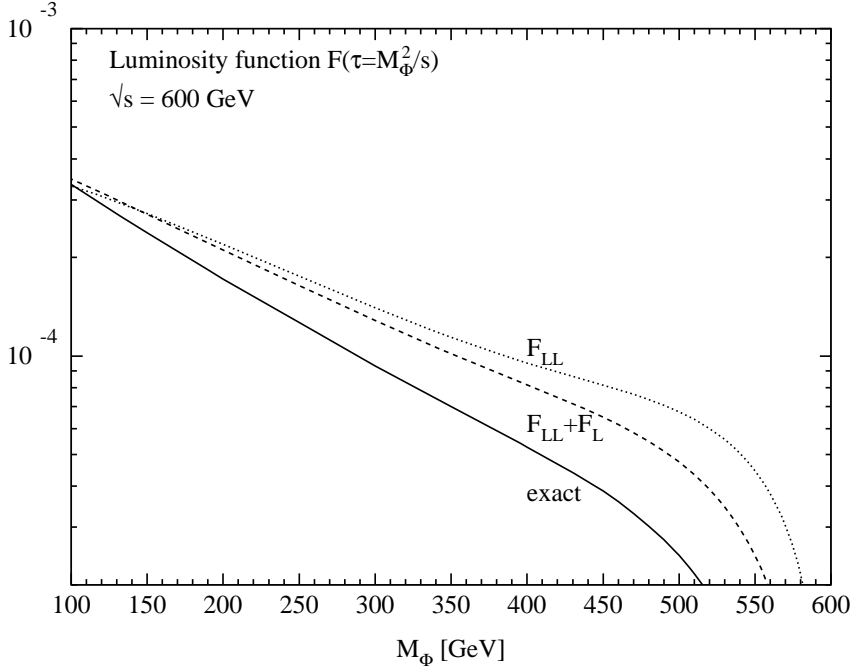,width=12cm}
\end{center}
\vskip -0.5cm
\caption{\it Comparison of the luminosity function $F$ calculated in the 
  leading-log approximation, $F_{LL}$, corrected with the single log,
  $F_{LL}+F_L$, and the exact tree-level calculation, presented as functions
  of the Higgs boson mass $M_\Phi$.}
\label{fig:ff}
\end{figure}

From these two elements the total $\gamma \gamma $ 
cross section can be calculated in the narrow-width approximation as
\begin{eqnarray}
   \sigma[\gamma \gamma \to \tau^+\tau^- \Phi ] 
        =\frac{\pi m_\tau^2}{2v^2 s} \, g^2_{\Phi\tau\tau} \times  
       2 \int_{\tau}^1 \frac{dz}{z}\,
          D^\tau_\gamma(z) D^\tau_\gamma(\tau / z )\,   
\label{gg2higgs}
\end{eqnarray}
where $\tau= M_\Phi^2/s$, with $s$ being the c.m.~energy squared 
of the photons.
Defining the $\tau\tau$ luminosity function 
by the convolution integral of the structure functions (including the
multiplicity factor 2), the leading-logarithmic part, $F_{LL}$, is given 
by
\begin{eqnarray}
  F_{LL}(\tau) &=&    
     \left(\frac{\alpha}{2\pi}\right)^2 \,
f_{LL}(\tau)\, \log^2 \frac{M^2_\Phi}{m^2_\tau}   \nonumber \\
    f_{LL}(\tau) & =&2 (1+2\tau)^2 \log\tau^{-1} -4(1-\tau)(1+3\tau)
  \label{LL}
\end{eqnarray}
The single log corrections can be included by replacing
$         F_{LL}  
     \to F_{LL}+F_L    
$
where\footnote{For a general factorization scale $\mu_F\neq M_\Phi$,  the term
  $2 f_{LL}\log\mu_F^2/ M^2_\Phi$ must be subtracted from $f_L$ in
  Eq.(\ref{FLtau}). }
\begin{eqnarray}
F_L(\tau) &=& \left(\frac{\alpha}{2\pi}\right)^2 f_L(\tau) 
\log  \frac{M^2_\Phi}{m^2_\tau} \nonumber \\
  f_L(\tau)  & =& 
  -8(1-\tau)(1+3\tau)\log(1-\tau) +(1+2\tau)^2[\log^2\tau
+4 Li_2(\tau)-\textstyle{\frac{2\pi^2}{3}}]\nonumber \\
&&          -(5+24\tau+4\tau^2)\log\tau^{-1}
      + \textstyle{\frac{1}{2}}(1-\tau)(27+103\tau)
  \label{FLtau}
\end{eqnarray}
Fig.~\ref{fig:ff} displays the luminosity function in the leading 
logarithmic approximation, $F_{LL}$, and corrected with the single log,
$F_{LL}+F_L$. Adding to the luminosity function the corrections from Higgs
bremsstrahlung off the external $\tau$ legs in the $\gamma\gamma$ process, the
generalized function $F(\tau)$, defined in parallel to the split form of 
Eq.(\ref{gg2higgs}) and calculated in the next section, 
is  shown here for comparison\footnote{This discussion has also been presented
in some detail as it sheds light on the quality of the 
parton picture for heavy 
quarks in high--energy proton beams where comparisons cannot be performed with 
the same rigor as in the lepton sector.}. Evidently, the analytical 
calculations provide us with a good-quality picture of the process so that a 
proper understanding of the mechanisms involved can be claimed.

For energies sufficiently above threshold, $F_{LL}$ provides a solid basis for
estimating the cross sections.  The analytical formula Eq.(\ref{LL}) can
readily be used therefore for a first estimate of the potential for measuring
$\tan\beta$ in the $b\bar{b}$-decay channel. A rough estimate shows the size
of the cross section to be $\sim 8$ fb for the $\gamma\gamma$ c.m. energy
$E_{\gamma \gamma} = 600$ GeV, $M_{H/A} = 400$ GeV and $\tan\beta = 30$. For
an integrated luminosity of 200 fb$^{-1}$, which may be accumulated in running
the $\gamma\gamma$ collider for one year, about 3000 events can be expected in
both $H$ and $A$ decay channels.  As a result, a statistical error of order
1\% can be predicted that compares favorably well with other methods 
\cite{R5,R6}. On the other hand the light Higgs boson $h$ and the heavy Higgs 
bosons $H,A$ for moderate mass values can also be produced at lower energies, 
{\it e.g.}~$E_{\gamma\gamma}=400$~GeV.

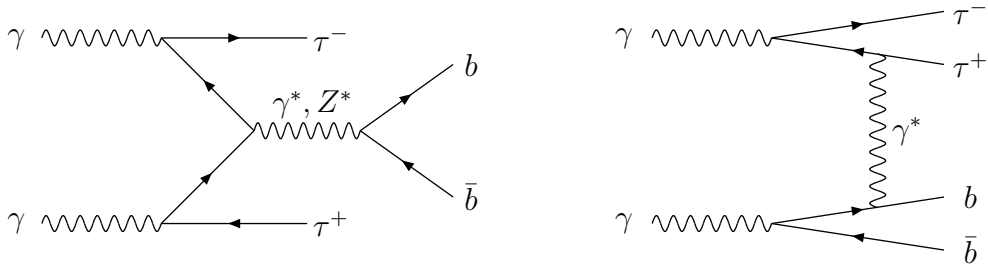
\begin{figure}[htb]
\begin{center}
\begin{picture}(700,100)(130,0)
\Text(171,85)[]{$\gamma$}
\Photon(180,85)(225,85){3}{7}
\Text(171,15)[]{$\gamma$}
\Photon(180,15)(225,15){3}{7}
\ArrowLine(260,50)(225,85)
\ArrowLine(225,15)(260,50)
\ArrowLine(225,85)(280,85)
\Text(290,85)[]{$\tau^-$}
\ArrowLine(280,15)(225,15)
\Text(290,15)[]{$\tau^+$}
\Photon(260,50)(300,50){3}{7}
\Text(283,60)[]{$\gamma^*,Z^*$}
\ArrowLine(335,25)(300,50)
\ArrowLine(300,50)(335,75)
\Text(343,75)[]{$b$}
\Text(343,25)[]{$\bar{b}$}
\Text(401,85)[]{$\gamma$}
\Photon(410,85)(455,85){3}{7}
\Text(401,15)[]{$\gamma$}
\Photon(410,15)(455,15){3}{7}
\ArrowLine(520,75)(455,85)
\ArrowLine(455,85)(520,95)
\Text(532,95)[]{$\tau^-$}
\Text(532,75)[]{$\tau^+$}
\ArrowLine(520,5)(455,15)
\ArrowLine(455,15)(520,25)
\Text(532,5)[]{$\bar{b}$}
\Text(532,25)[]{$b$}
\Photon(495,21)(495,79){3}{9}
\Text(508,50)[]{$\gamma^*$}
\end{picture}
\end{center}
\caption{\it The annihilation and diffractive 
  background processes in $\gamma\gamma$ collisions giving rise to
  $\tau\tau{b}b$ final states. \label{fig:bkgd}}
\end{figure}

In the same way we can estimate the size of the cross section for the main
background channel: $\tau^+ \tau^-$ annihilation into a pair of $b$-quarks,
$\tau^+ \tau^- \to b \bar{b}$, via $s$-channel $\gamma$ and $Z$ exchanges,
Fig.~\ref{fig:bkgd} left panel. As the mechanism is electroweak, the cross
section is naturally small (what would have been different for 4$b$ final
states where strong QCD processes would be activated). If proper care is taken
for the invariant mass of any of the fermion pairs not to match the $Z$-boson
mass (except for $h$), the transition probability is reduced by one power of
the electroweak coupling squared compared with the signal process. The $\gamma
\gamma$ cross section of this background channel is obtained by integrating
the parton cross section $\hat{\sigma}[\tau^+\tau^- \to \gamma, Z \to b\bar{b}]
 = 4\pi\alpha^2 \left < |Q_b Q_\tau|^2 \right >/\hat{s}$, 
with the familiar generalized electroweak charges $Q_\tau$ and $Q_b$, 
over the $\tau\tau$ luminosity of the $\gamma \gamma$ collider within the
range $\Delta$ at the invariant energy $\sqrt{\hat{s}} = M_\Phi$. 
The integration range 
$\Delta$ is taken either as the total width of the Higgs bosons $\Gamma_\Phi$ 
or as the estimated experimental resolution $\Delta_{ex} \simeq \pm 0.05 \, 
M_\Phi$ of the $b\bar{b}$ final state \cite{R8}:
\begin{eqnarray}
   \Delta = \mathrm{max}[\Gamma_\Phi, \, 2\Delta_{ex}]      \label{deltaM}
\end{eqnarray} 
Inserting typical parameters for a Higgs boson mass of $M_{H/A} = 400$ GeV,
one finds a background cross section of size $\lsim 5\times 10^{-3}$ fb.  This
value, corresponding roughly to the signal for $\tan\beta\sim 2$, is much
smaller than the size of the signal cross section for large values of
$\tan\beta \gsim 10$.  Similar signal-to-background ratios are predicted for
the light Higgs boson $h$, except for masses close to the $Z$-boson mass.  For
small Higgs-boson masses, the background gradually increases when the
$b\bar{b}$ invariant mass approaches the on-shell $Z$-boson mass and the
background suppression as a result of the additional electroweak coupling
ceases to be effective.

The topology of the related background process of $b\bar{b}$ fusion to a
$\tau^+\tau^-$ pair via $\gamma$ or $Z$ exchange is quite different from the
signal. It can be suppressed to a very small level in two ways: by requiring
sufficiently large transverse momenta of the $b$ quarks and
sufficiently small transverse momenta for the $\tau$'s.

A second background channel is associated with diffractive $\gamma \gamma \to
(\tau^+ \tau^-) (b\bar{b})$ events, the pairs scattering off each other by
Rutherford photon exchange, Fig.~\ref{fig:bkgd} right panel. While being very
large in principle, the diffractive background can be suppressed strongly by
proper cuts. The paired fermions in diffractive events travel preferentially
parallel to the $\gamma$ axes and they carry small invariant mass. Requiring
therefore a large invariant mass for the $b\bar{b}$ pair, equivalent to the
Higgs signal, and, as suggested by the topology of the signal, the $\tau$'s to
go into opposite directions near the beam axis, the background can be reduced
strongly. This expectation is borne out by the quantitative numerical analysis
presented in the
next section.\\

\vskip 3mm
\noindent
{\bf 3. \underline{Numerical Analysis.}}
Encouraged by the semi-quantitative estimates, we have
performed a detailed numerical analysis for signal and backgrounds.

Assuming an $e^\pm e^-$ collider c.m. energy of 800 GeV, the maximum of $\gamma
\gamma$ energy spectrum 
can be taken as 600 GeV. Adopting the TESLA parameters, for
instance, an integrated $\gamma\gamma$ 
luminosity of about 200 fb$^{-1}$ {\it per annum} can
be expected in the margin 20\% below the maximum $e^+e^-$ 
energy \cite{R10}. Similarly,
about 100 fb$^{-1}$ may be accumulated for a
$\gamma\gamma$ energy of 400 GeV at a 500 GeV $e^\pm e^-$ collider. 

\begin{figure}[ht!]
\begin{center}
\epsfig{figure=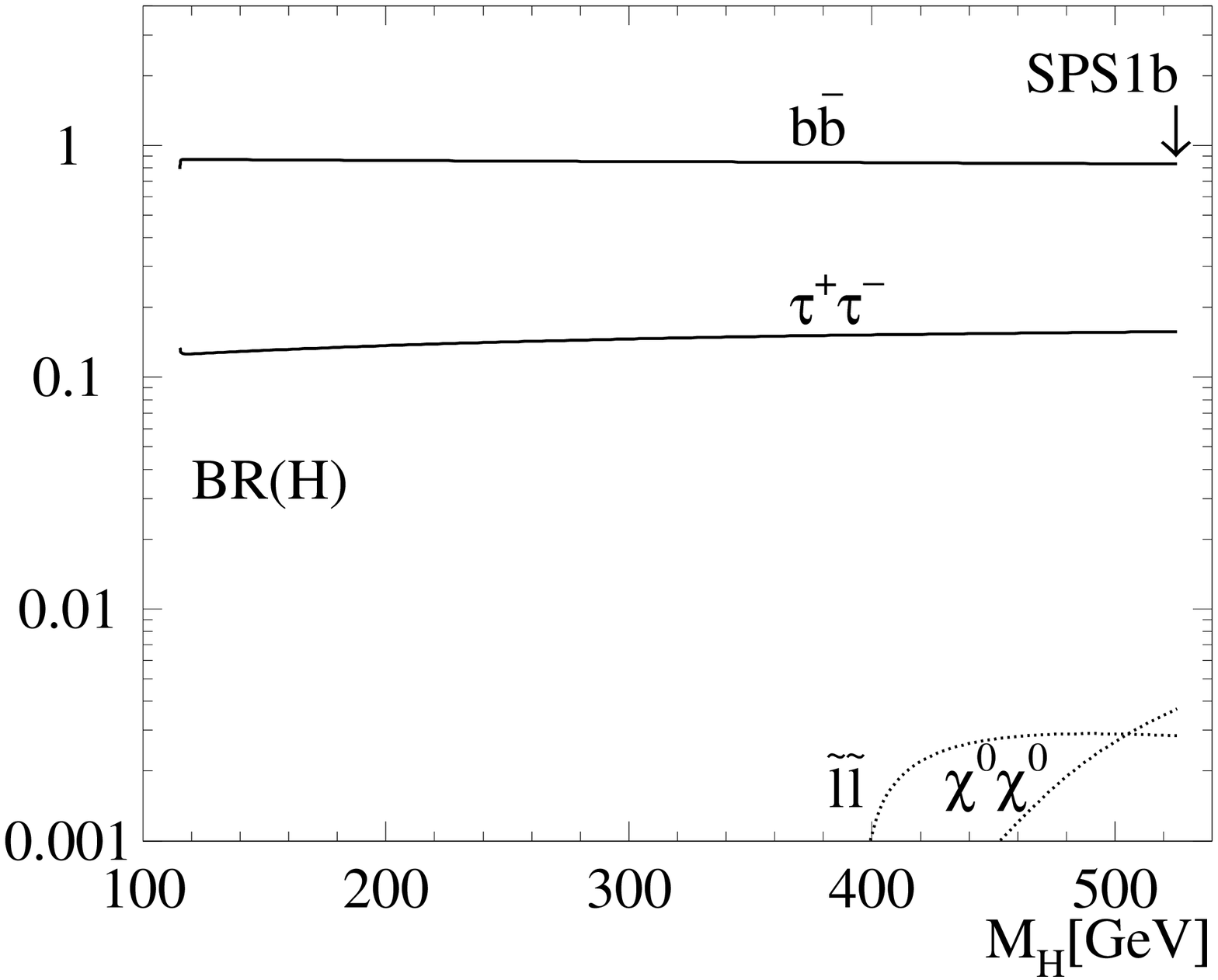,width=7.5cm}
\epsfig{figure=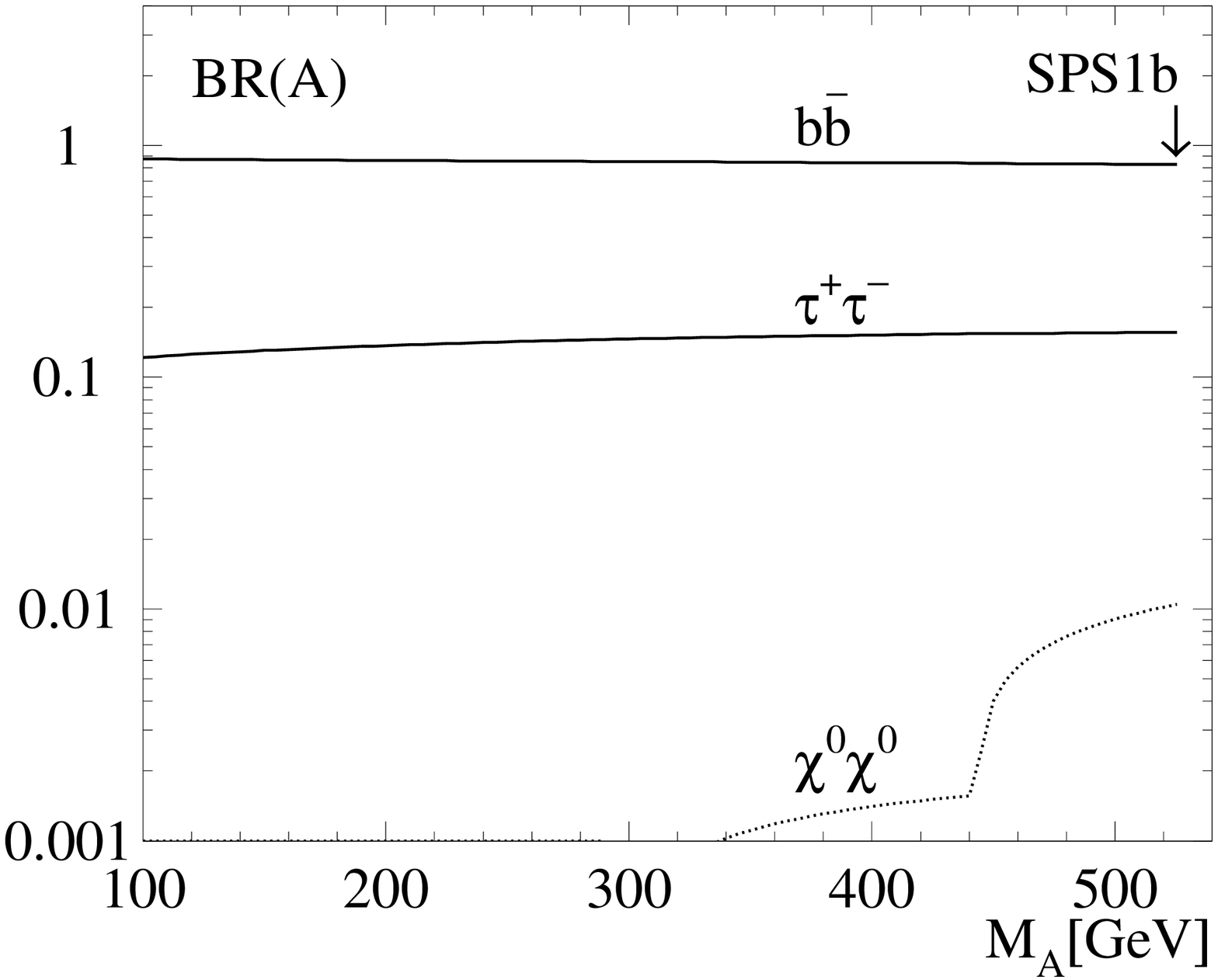,width=7.5cm}\\
\epsfig{figure=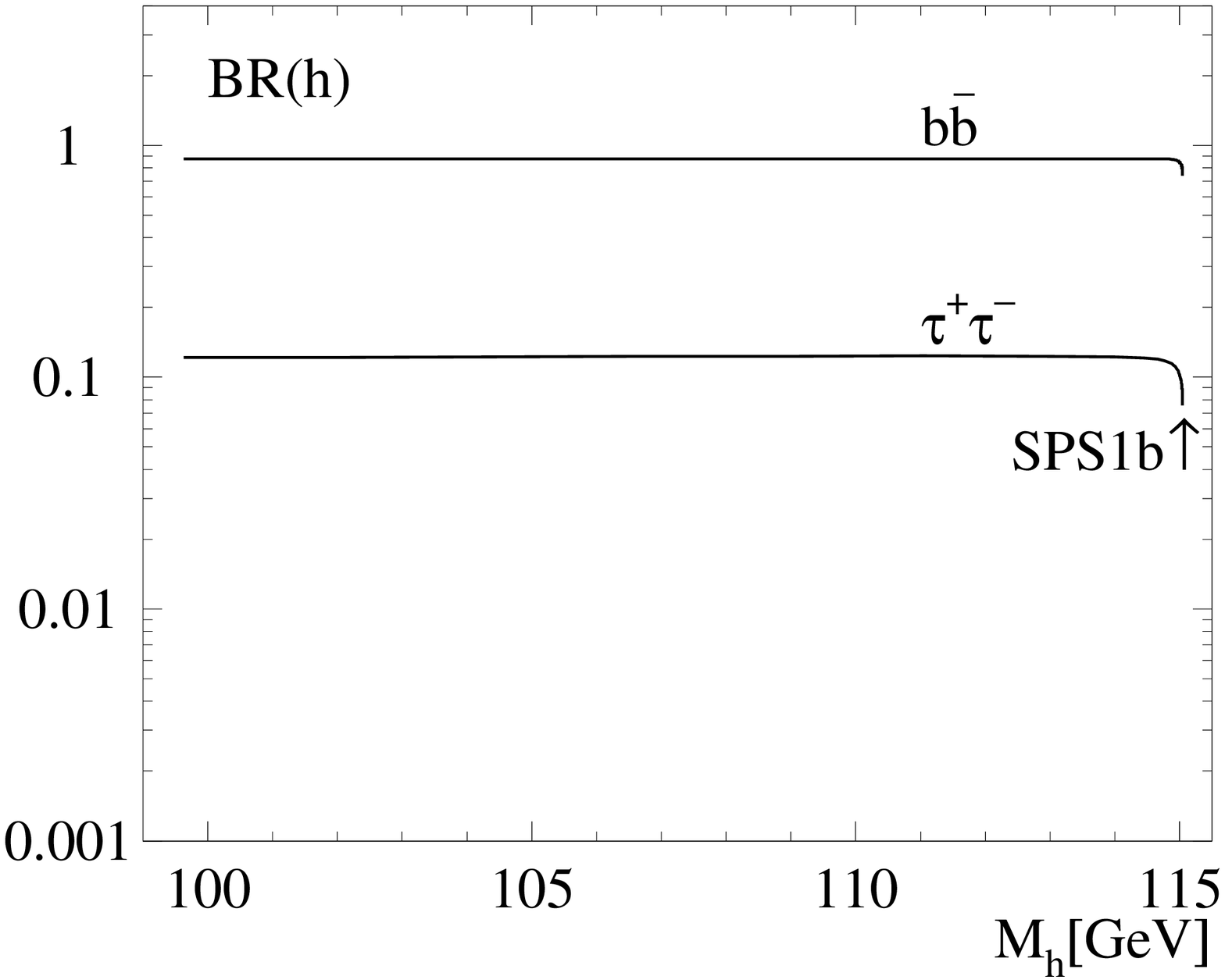,width=7.5cm}
\epsfig{figure=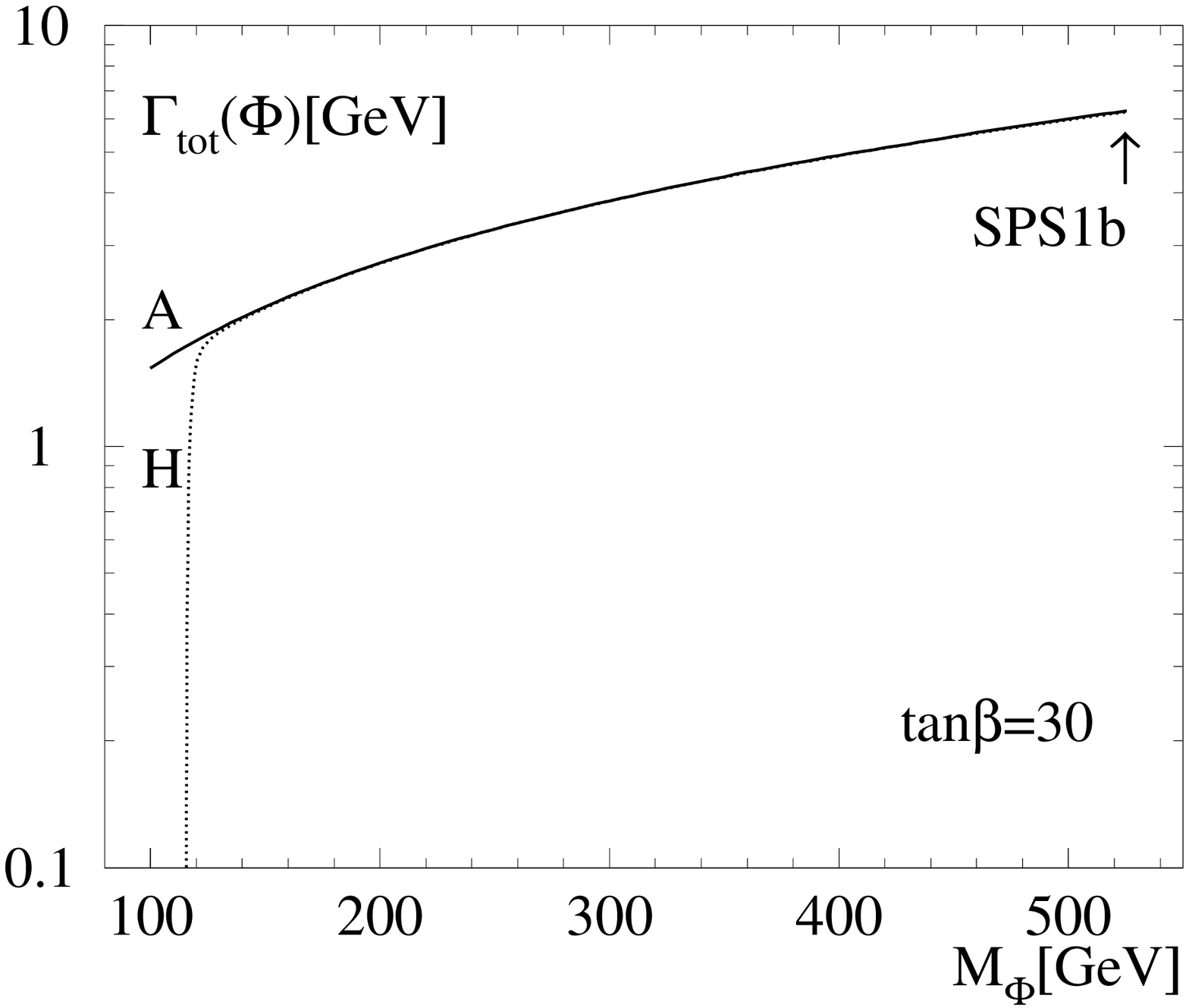,width=7.5cm}
\end{center}
\vskip -0.5cm
\caption{\it The decay  branching ratios of the Higgs bosons 
  $H$ and $A$ (upper row) and $h$ (lower left). The lower right panel shows
  the total decay widths of $H$ and $A$. All quantities are given as functions
  of the respective Higgs boson mass.  The SUSY parameters, in particular
  $\tan\beta=30$, have been chosen as in the SPS1b scenario except for $M_A$
  which varies from 100~GeV to 525~GeV. The upper limit corresponds to the
  SPS1b Snowmass point.}
\label{fig:BR}
\end{figure}

As alluded to before, we shall analyze the system under the assumption that
Higgs decay channels to supersymmetric particles are shut. Such a scenario is
realized approximately in the Snowmass reference point SPS1b \cite{R11} for
which decays to charginos and neutralinos add up to a branching ratio of less
than 1\%, see Fig.~\ref{fig:BR}. In this situation the $H/A$ Higgs bosons
decay primarily to $bb$ pairs with a small admixture of $\tau\tau$
pairs for large $\tan\beta$: $BR_{\tau}/BR_b = 1/3 \, (m_{\tau}/m_b)^2 \sim
0.1 $ for Higgs masses of 400 GeV. The $H/A$ decay widths, which are important
parameters to control the background suppression, are also shown in
Fig.~\ref{fig:BR} for $\tan\beta = 30$ as a function of the Higgs masses. In
the mass range of interest the total Higgs decay widths are of the order of a
few GeV.  Moreover, they are comparable to the experimental $b\bar{b}$
invariant mass resolution, and the resonant $b$-quark jets from the Higgs
boson decay should clearly be visible above the smooth background. The light
Higgs boson $h$ has similar decay branching ratios for moderately small $M_A$.

\begin{figure}[ht!]
\begin{center}
\epsfig{figure=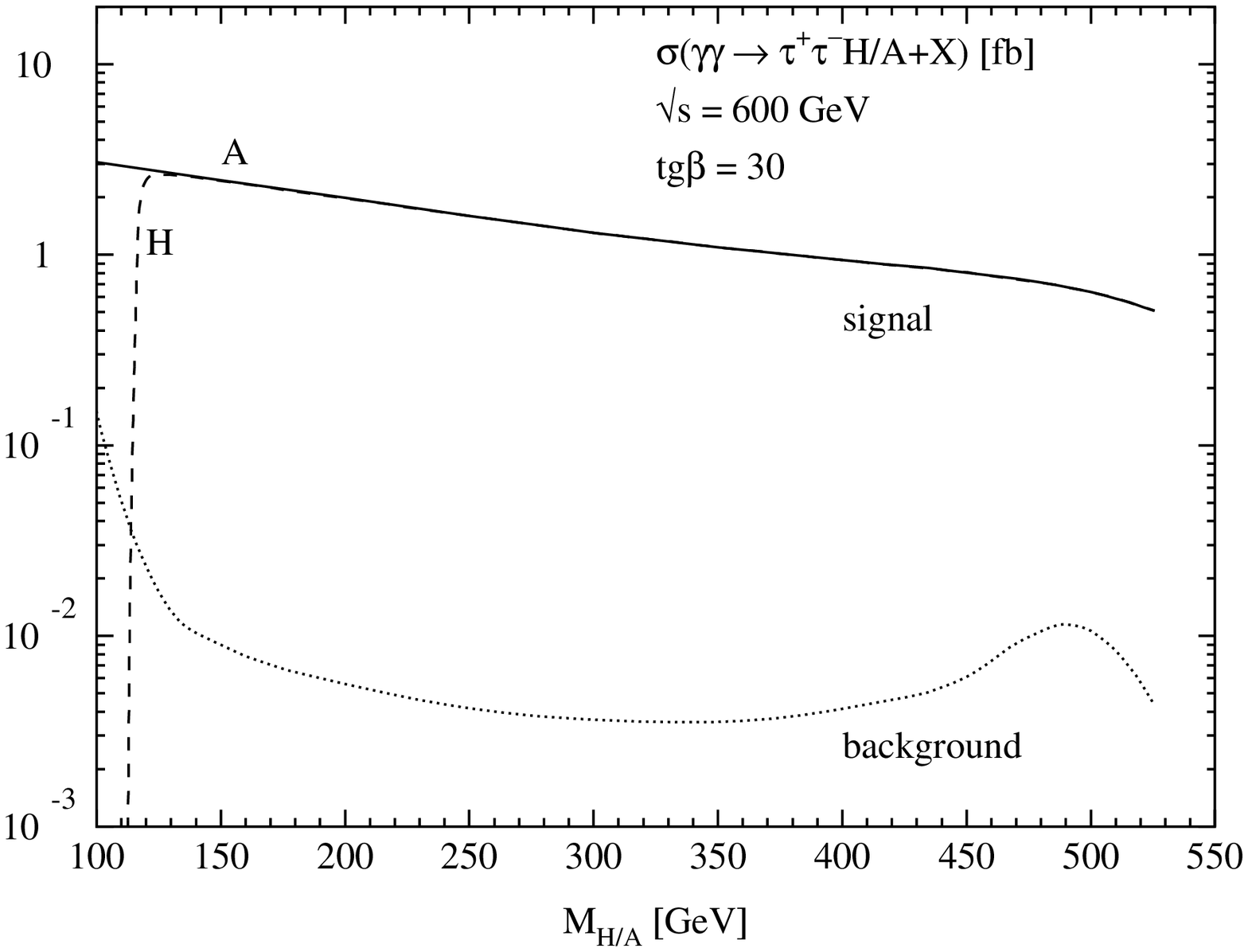,height=9cm}\\
\epsfig{figure=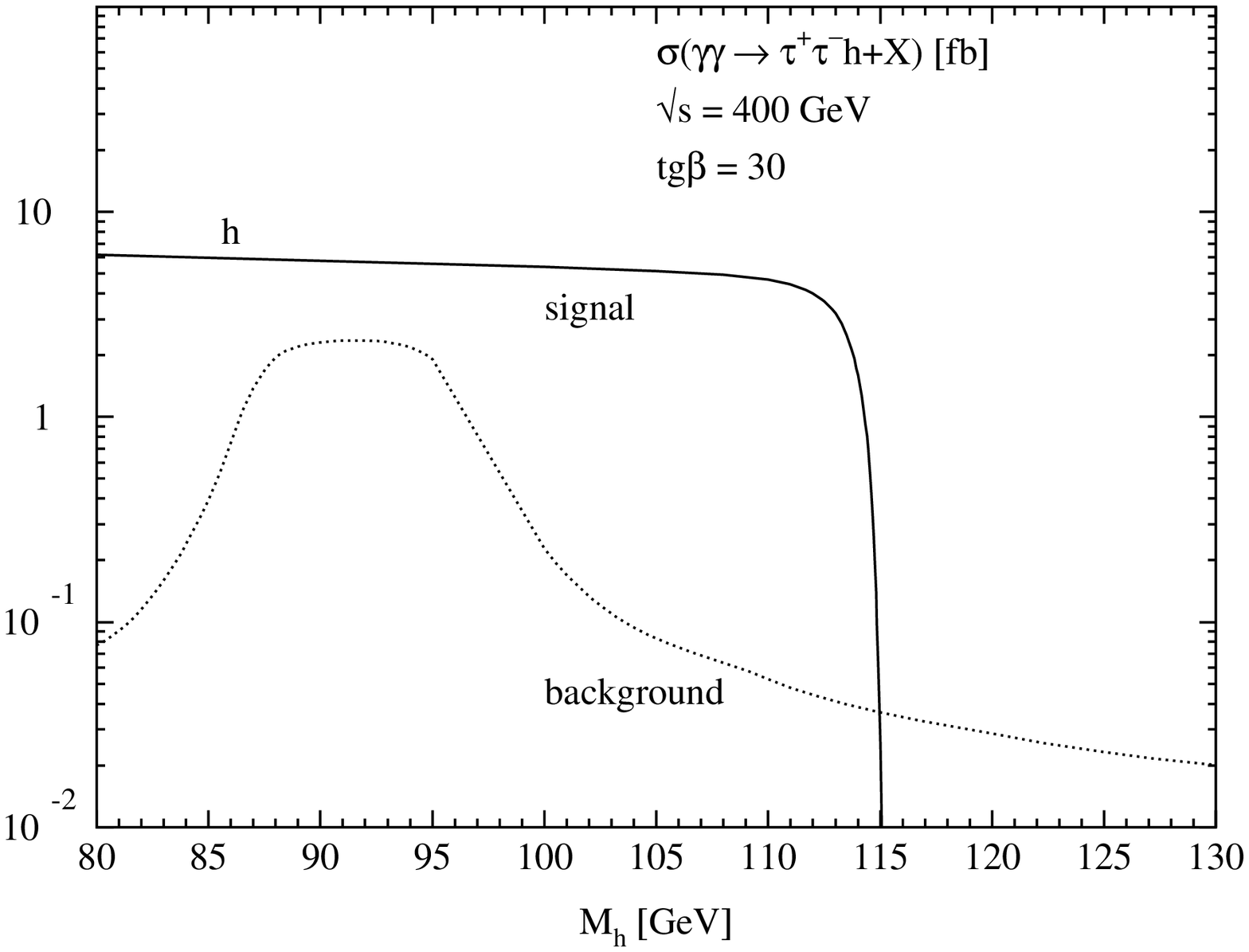,height=9cm}
\end{center}
\vskip -0.5cm
\caption{\it The cross sections for the production of the $H/A$ (top) 
  and $h$ (bottom) Higgs bosons in the $\tau\tau$ fusion process for
  $\tan\beta=30$.  Also shown is the background cross section for experimental
  cuts as specified in the text.  SUSY parameters as in Fig.\ref{fig:BR}.
  $\sqrt{s}$ denotes the $\gamma\gamma$ collider c.m.~energy, corresponding to
  approximately 80\% of the $e^\pm e^-$ linear collider energy.}
\label{fig:xsec}
\end{figure}
In the numerical analysis the full set of diagrams for the signal processes
$\gamma \gamma \to \tau^+ \tau^- + \Phi [\to b\bar{b}]$ and all diagrams 
for the background processes $\gamma \gamma \to \tau^+ \tau^- b\bar{b}$, 
generated by means of {\tt CompHEP} \cite{R12}, are taken into account.
This set includes for the signal in particular bremsstrahlung of the
Higgs bosons off the external $\tau$ legs in the non-logarithmic
corrections to the equivalent-particle approximation.
  
We finally present in the top panel of Fig.~\ref{fig:xsec} the exact cross
sections for the signals coming from the production of $H$ and $A$ Higgs
bosons in the $\tau\tau$ fusion process together with all the background processes.
The cuts which define the final states have been chosen such that the
diffractive $\gamma$-exchange mechanism is suppressed sufficiently well.  For
this purpose, the invariant $b\bar{b}$ mass has been constrained to the
bracket $ \Delta$ as defined in Eq.(\ref{deltaM})\footnote{On the $M_{A,H}$
  slope of SPS1b analyzed here in detail, the heavy Higgs bosons remain so
  narrow that $\Delta$ is always given by the experimental resolution
  $\Delta_{ex}$.}, the taus are assumed visible and traveling in opposite
directions to the beam axis with tau energies in excess of 5 GeV and polar 
angles beyond 130 mrad to account for the shielding.
Note that no cut on the $\tau^+\tau^-$ invariant mass is
required to suppress the $b\bar{b}$ fusion process to $Z$ bosons decaying to
$\tau$ pairs which increases the background cross section marginally for large
Higgs masses before it is cut off completely by phase space.  Even $\tan\beta$
values down to $\tan\beta \gsim 5$ could be probed without running into
background problems. [The discussion of experimental cuts on transverse 
momenta and total event energy to remove overlaying $\gamma\gamma$ events is 
beyond the scope of this theoretical study.] 

In the complementary bottom panel of Figure \ref{fig:xsec} it is shown that
also $\tau\tau$ fusion to the light Higgs boson $h$ can be exploited to measure
$\tan\beta$ for large values if the pseudoscalar mass is moderately small.  In
the example presented in Fig.~\ref{fig:xsec}, the $\gamma \gamma$ c.m.~energy
has been set to 400 GeV.

The drop in the $H$ and $h$ signal cross sections at the lower and upper end
of the mass range, respectively, is caused by the 
reduction of the $H\tau\tau$ and
$h\tau\tau$ couplings to the Standard Model values in these parameter areas. \\

\begin{table}[ht!]
\begin{center}
\begin{tabular}{|c||c|cc||c|cccc|}
\hline\rule{0cm}{4mm}
 &\multicolumn{3}{|c||}{$E_{\gamma\gamma}=400$ GeV, ${\cal L}=100$ fb$^{-1}$} 
 &\multicolumn{5}{|c|}{$E_{\gamma\gamma}=600$ GeV, ${\cal L}=200$ fb$^{-1}$}\\
\hline
\rule{0cm}{5mm}$M_{\mathrm{Higgs}}$ & $A\oplus h$ &
\multicolumn{2}{|c||}{$A\oplus H$}& 
$A\oplus h$ &\multicolumn{4}{|c|}{$A\oplus H$}\\ 
 \phantom{i}[GeV]  
&  100  &  200  &  300  & 100   &  200  &  300  &  400  &  500  \\
\hline
\rule{0cm}{5mm}$\tan\beta$ & \scriptsize{I}   & \scriptsize{II}   & 
              \scriptsize{III} &
              \scriptsize{IV}  & \scriptsize{V}    & \scriptsize{VI}  &
              \scriptsize{VII} & \scriptsize{VIII}     \\
\hline\hline
\rule{0cm}{5mm}10  & 8.4\% & 10.7\% & 13.9\% & 8.0\% & 9.0\% & 
11.2\% & 13.2\% & 16.5\% \\
30  & 2.6\% & 3.5\% & 4.6\% & 2.4\% & 3.0\% & 3.7\% & 4.4\% & 5.3\% \\
50  & 1.5\% & 2.1\% & 2.7\% & 1.5\% & 1.8\% & 2.2\% & 2.6\% & 3.2\% \\
\hline
\end{tabular}
\caption{\it Relative errors $\Delta\tan\beta/\tan\beta$ on $\tan\beta$ 
measurements for $\tan\beta=$ 10, 30 and 50 based on: 
combined $A\oplus h$ [I,IV] and $A\oplus H$ [II,III,V--VIII] production, 
assuming $E_{\gamma\gamma}= 400$~GeV, ${\cal L}= 100$~fb$^{-1}$
and  $E_{\gamma\gamma}= 600$~GeV, 
${\cal L}= 200$~fb$^{-1}$. Cuts and efficiencies are applied on the 
final--state $\tau$'s and $b$ jets as specified in the text.}
\label{tab:stat-error}
\end{center} 
\end{table}

The statistical accuracy with which large $\tan\beta$ values can be measured in
$\tau\tau$ fusion to Higgs bosons can be estimated from the predicted cross
sections and the assumed integrated luminosities. Efficiencies for $bb$ 
tagging, $\epsilon_{bb}$, and $\tau\tau$ tagging, $\epsilon_{\tau\tau}$, 
reduce the accuracies. For $\epsilon_{bb}\sim 0.7$ and 
$\epsilon_{\tau\tau}\sim 0.5$, for example \cite{R8}, the errors grow by a 
factor $1/\sqrt{ \epsilon_{bb}\epsilon_{\tau\tau}}\sim 1.7$. The expected 
errors for $h/H/A$ production are exemplified for three 
$\tan\beta$ values, $\tan\beta=10$, 30 and 50, in Table~\ref{tab:stat-error}. 
The integrated luminosities are chosen to be 200 fb$^{-1}$ for the high 
energy option and 100 fb$^{-1}$ for the low energy option \cite{R10}.
For $h$ production, the mass parameters are set to $M_A\sim 100$ GeV and
$M_h=100$ GeV; for the production of the heavy pseudoscalar $A$ the mass is
varied between 100 and 500 GeV. Results for scalar $H$ production are 
identical to $A$ in the mass range above 120 GeV. The two channels $h$ and 
$A$, and $H$ and $A$ are combined in the overlapping mass ranges in which
the two states, respectively, cannot be discriminated. In 
Table~\ref{tab:stat-error} we have presented the relative errors 
$\Delta\tan\beta/\tan\beta$. Since in the region of interest 
the $\tau\tau$ fusion cross sections are proportional to $\tan^2\beta$ and 
the background is small, the absolute errors $\Delta\tan\beta$ are nearly 
independent of $\tan\beta$, varying between 
\begin{eqnarray}
\Delta\tan\beta \simeq 0.9 {\;\;\;\rm and \;\;\;} 1.3          \label{deltat}
\end{eqnarray}
for Higgs mass values away from the kinematical limits.

It should be noted that away from the kinematical limits, the Higgs fusion
cross sections $\sim F/s$, Eq.(\ref{gg2higgs}), vary little with the
$\gamma\gamma$ energy since the suppression by $s$ is almost compensated by
the luminosity function $F$.  As a result, the smearing of the
$\gamma\gamma$ energy has a mild effect on the analysis presented here. 
Moreover, since the $\gamma\gamma$ luminosity rises with the collider energy,
the errors on $\tan\beta$ decrease.  Of course, detailed experimental
simulations are required for the final conclusions on the anticipated
accuracies.  However, the above estimates can clearly be interpreted as
encouraging experimental simulations with promising prospects for valuable
measurements of $\tan\beta$.

\vskip 3mm
\noindent
{\bf 4. \underline{Summary.}}  
We have demonstrated in this letter that $\tau\tau$ fusion to the heavy Higgs
bosons $H/A$ of the MSSM at a photon collider is a promising channel for
measuring the Higgs mixing parameter $\tan\beta$ at large values. Complemented
by $\tau\tau$ fusion to the light Higgs boson $h$ for moderately small values 
of the pseudoscalar Higgs boson mass $M_A$, the MSSM parameter range can nicely
be covered in all scenarios.  The analysis compares favorably well with the
corresponding $b$-quark fusion process at the LHC \cite{R6}.  Moreover, the 
method can be applied readily for a large range of Higgs mass values and thus 
is competitive with complementary methods in the $e^+e^-$ mode of a linear 
collider \cite{R5}.\\

\noindent {\bf  Acknowledgments:} We thank K. Desch and V. Telnov for
valuable advise on experimental parameters relevant for this study.
Remarks on the manuscript by K.~Desch, M.~Krawczyk and 
D.~Rainwater are gratefully acknowledged. 

The work is supported in part by the European Commission 5-th Framework
Contract HPRN-CT-2000-00149. The work of SYC was supported in part by the
Korea Research Foundation Grant (KRF--2002--070--C00022) and in part by KOSEF
through CHEP at Kyungpook National University. JK was supported by the KBN
Grant 2 P03B 040 24 (2003-2005) and 115/E-343/SPB/DESY/P-03/DWM517/2003-2005.
The work of JSL was supported in part by PPARC. MMM and MS have been supported
in part by the Swiss Bundesamt f\"ur Bildung und Wissenschaft.


\end{document}